
\magnification=1200\baselineskip=18pt
{\nopagenumbers
\centerline{\bf Dyonic black holes in effective string theory}
\vskip40pt
\centerline{\bf S. Mignemi}
\vskip20pt
\centerline {Dipartimento di Scienze Fisiche, Universit\`a di Cagliari}
\centerline{via Ospedale 72, 09100 Cagliari, Italy}
\vskip40pt
{\noindent The spherical symmetric dyonic black hole solutions of the
effective action of heterotic string are studied perturbatively up to
second order in the inverse string tension. An expression for the
temperature in terms of the mass and the electric and magnetic charge
of the black hole is derived and it is shown that its behaviour is
qualitatively different in the two special cases where the
electric or the magnetic charge vanishes.}
\vskip100pt
\centerline{February 1993}
\vfil\eject}
\def\lapp{[(r-2m)}
\def\alp{\alpha '}\def\GB{Gauss-Bonnet }\def\Sch{Schwarzschild }
\def\RN{Reissner-Nordstr\"om }
\def\ri{{\cal R}}\def\gb{{\cal S}}\def\ef{{\rm e}^{-2\Phi}}
\def\fo{\phi_1}\def\ft{\phi_2}\def\ro{\rho_1}\def\rt{\rho_2}
\def\po{\psi_1}\def\pt{\psi_2}\def\co{\chi_1}\def\ct{\chi_2}

\def\ep{\epsilon_{abcd}}\def\e{{\rm e}}
\def\lap{[r(r-2m)}\def\bc{boundary conditions }

At present, string theories are the most promising attempt to unify
gravitation
with the other fundamental interactions. It is therefore of great
interest to study their phenomenological consequences, in particular in the
context of gravitational physics.

For this reason, the modifications to black hole physics induced by string
theory have been extensively investigated. Due to the difficulty of treating
the full theory,
a fruitful approach has been the study of its low energy limit by means
of effective field theoretic actions, obtained as an expansion in
the inverse string tension $\alpha '$. These actions describe the dynamics
of the low energy excitations of the string spectrum, which comprise in
the simplest case the graviton, a dilaton, an axion and gauge fields.
In particular, some exact solutions have been obtained for the lowest order
approximation to the action [1-3]. The main difference from the Einstein
theory is the presence of a non-minimally coupling of the dilaton with the
other fields, which changes drastically the physical properties of the black
hole, and in particular the thermodynamics. For charged black holes, for
example, the temperature becomes independent of the charge, at variance with
the \RN solution of the Einstein-Maxwell theory.

The results obtained in [1-3] are however valid only if the order $\alpha '$
corrections in the gravitational sector of the effective action are neglected.
These consist essentially in the presence of a higher derivative \GB term
coupled to the dilaton.
In a recent paper [4], we have examined how the magnetic charged dilatonic
black hole solutions of effective string theories obtained in [1] are modified
when the order $\alpha '$
terms in the effective action are taken into account. Of particular interest
were the thermodynamical properties of the solutions, which, contrary to the
solutions of the leading order action, exhibit a dependence of the temperature
on the magnetic charge of the black hole. In particular, these solutions can
reach a vanishing temperature
for a finite value of the mass, and therefore admit a stable remnant as
final state of the Hawking evaporation.

In this paper we extend our investigation to the case of dyonic black holes.
When higher order corrections to the action are neglected, the dyonic solution
has been obtained by the magnetic one [1] by exploiting the SL(2,R)
dilaton-axion symmetry of the leading order terms in the perturbative
expansion of the action [2]. Even if it has been conjectured that this
symmetry might be an exact symmetry of the full theory [3],
it is not however a symmetry of the order $\alpha '$ action, since the \GB
term in that action is not invariant under the $SL(2,R)$ symmetry. It has been
argued however that the symmetry could still be present in a highly
non-trivial and non-linear way and therefore not order by order in
perturbation theory [5].

For these reasons we perform here a perturbative calculation of the
spherical symmetric solutions of the order $\alpha '$ action around the
background costituted by the \Sch solution.  The most interesting result of
our analysis is the difference in the behaviour of the temperature for an
electrically  charged black hole with respect to the magnetic one when
higher order terms are taken into account. In fact, it appears that the
corrections to the temperature have opposite sign in the two cases, so that
for a purely electric black hole in this theory the temperature is a
monotonic decreasing positive definite function of the mass for any value
of the charge and therefore does not give rise to stable remnants.

The bosonic sector of the dimensionally reduced effective action for the
heterotic string is given up to order $\alp$ by [6]:
$$S_{eff}=\int d^4 x \sqrt{-g}\left[\ri-{1\over 3}\e^{-4\Phi}H_{abc}H_{abc}
-2(\nabla\Phi)^2+\alpha\ef(\gb-F^2)\right]
\eqno(1)$$
where $\alpha\equiv\alp /8$, $F$ is the Maxwell field strength,
$\gb\equiv\ri^2_{abcd}-4\ri^2_{ab}+\ri^2$ is
the Gauss-Bonnet term and we have omitted the terms which do not contribute
to the field equations up to order $\alpha^2$.
The axion field $H_{abc}$ is given explicitly in terms of the potential
$B_{ab}$ as:
$$H_{abc}=\partial_{[a} B_{bc]}+\alpha(A_{[a}\partial_b A_{c]}+\omega^{lm}_
{\quad[a}\partial_b\omega^{lm}_{\quad c]}+\omega^{lm}_{\quad [a}\omega^{mn}_
{\quad b}\omega^{nl}_{\quad c]}),$$
where the Chern-Simons terms have been taken into account. Due to the
presence of these terms
a solution presenting both electric and magnetic charge will also have a
non-trivial axion field [6].

It is well known that if one defines a scalar field $a$, dual to $H$, such
that
$$H_{abc}=-{1\over 2}\ep\e^{4\Phi}\partial_d a
\eqno(2)$$
the field equations can be derived from the action:
$$S_{eff}=\int d^4 x \sqrt{-g}\left[\ri-2(\nabla\Phi)^2
-{1\over2}\e^{4\Phi}(\nabla a)^2-\alpha\ef(F^2-\gb)\right]-\alpha a(F\tilde F
-\ri\tilde\ri)
\eqno(3)$$
where
$$F\tilde F ={1\over 2}\ep F_{ab}F_{cd}\qquad \ri\tilde\ri ={1\over 2}\ep
\ri_{abef}\ri_{cdef}
\eqno(4)$$
The field equations can then be written as:
$$\eqalign{\ri_{mn}=&2\nabla_m\Phi\nabla_n\Phi+{1\over 2}\nabla_m a\nabla_n a
+2\alpha\ef (F_{mp}F_{np}-{1\over 4}
g_{mn}F^2)\cr
&+4\alpha\ef\Big[ 4\ri_{p(m}\nabla_{n)}\nabla_p\Phi-2\ri_{mn}\nabla_p\nabla_p
\Phi-g_{mn}\ri_{pq}\nabla_p\nabla_q\Phi\cr
&-\ri\nabla_m\nabla_n\Phi+{1\over2}
g_{mn}\ri\nabla_p\nabla_p\Phi-2\ri_{qmnp}\nabla_p\nabla_q\Phi\Big]\cr
&-8\alpha\ef\Big[ 4\ri_{p(m}\nabla_{n)}\Phi\nabla_p\Phi-2\ri_{mn}\nabla_p\Phi
\nabla_p\Phi-g_{mn}\ri_{pq}\nabla_p\Phi\nabla_q\Phi\cr
&-\ri\nabla_m\Phi\nabla_n
\Phi+{1\over2}
g_{mn}\ri\nabla_p\Phi\nabla_p\Phi-2\ri_{qmnp}\nabla_p\Phi\nabla_q\Phi\Big]\cr
&+4\alpha\Big[\epsilon_{blmn}\ri_{ajmn}\nabla_l\nabla_j a+
\epsilon_{almn}\ri_{bjmn}\nabla_l\nabla_j a\Big]\cr}
\eqno(5a)$$
$$\nabla^2\Phi={\alpha\over 2}\ef (\gb -F^2)+{1\over2}\e^{4\Phi}(\nabla_p a)^2
\eqno(5b)$$
$$\nabla_p(\e^{4\Phi}\nabla_p a)=\alpha(F\tilde F-\ri\tilde\ri)
\eqno(5c)$$
$$\nabla_p(\ef F_{pm})={1\over 2}\epsilon_{pmqr}F_{qr}\nabla_r a
\eqno(5d)$$
It is convenient to define a new scalar field $b$, such that
$$\nabla_p b=\e^{4\Phi}\nabla_p a
\eqno(6)$$
The field equation can then easily be written in term of $b$. In particular:
$$\nabla^2\Phi={\alpha\over 2}\ef (\gb -F^2)+{1\over2}\e^{-4\Phi}(\nabla b)^2
\eqno(7a)$$
$$\nabla^2 b=\alpha(F\tilde F-\ri\tilde\ri)
\eqno(7b)$$

We want to find a perturbative expansion of the solution to (5)
around the background constituted by the Schwarzschild metric of mass $m$
with vanishing dilaton and axion. Our expansion
will be in the parameter $\alpha$ or, more correctly, in ${\alpha\over m^2}$.
We adopt a spherically symmetric ansatz for the metric:
$$ds^2=-\lambda^2 dt^2+\lambda^{-2}dr^2+R^2d\Omega^2
\eqno(8)$$
where $\lambda=\lambda(r)$, $R=R(r)$. This particular form of the metric is
suggested from the exact solution found in [2] in the absence of the
\GB term (see also [4]).

The general spherical symmetric solution of the generalized Maxwell equation
(5d) containing both electric and magnetic charge $q_e$ and $q_m$
respectively, is given in orthonormal
coordinates by:
$$F_{ij}={q_m\over R^2}\qquad\qquad F_{01}={q_e+q_m a\over R^2}\e^{2\Phi}
\eqno(9)$$
We can now expand the fields in $\alpha$ as:
$$\lambda=\lambda_0(1+\alpha\po +\alpha^2\pt+\dots) \qquad\qquad
R=r+\alpha\ro+\alpha^2\rt+\dots$$
$$\Phi=\alpha\fo+\alpha^2\ft+\dots \qquad\qquad b=\alpha\chi_1+\alpha^2\chi_2
+\dots
\eqno(10)$$
where $\lambda_0=(1-{2m\over r})^{1/2}$ and $\psi_i$, $\rho_i$, $\phi_i$ and
$\chi_i$ are functions of $r$.
Moreover, from (9)
$$F_{ij}\sim{q_m\over r^2}\left(1-{2\alpha\ro\over r}+\dots\right)
\qquad\qquad F_{01}\sim{q_e\over r^2}+\alpha\left[{q_e\over r^2}\left(2\fo
-{2\ro\over r}\right)+{q_m\over r^2}\co\right]+\dots
\eqno(11)$$
and hence:
$$F^2\sim 2{q_m^2-q_e^2\over r^4}+2\alpha\left[-4q_m^2{\ro\over r^5}
+4q_e^2\left({\ro\over r^5}-{\fo\over r^4}\right)-2q_eq_m{\co\over r^4}\right]
+\dots
\eqno(12a)$$
$$F\tilde F\sim 4{q_e q_m\over r^4}+4\alpha\left[q_m^2{\co\over r^4}
+2q_eq_m\left({\fo\over r^4}-{2\ro\over r^5}\right)\right]+\dots
\eqno(12b)$$

The field equations can now be expanded by inserting (10) and (11). At first
order in $\alpha$ one has:
$$\lap\fo ']'={q_e^2-q_m^2\over r^2}+{24m^2\over r^4}
\eqno(13a)$$
$$\lap\co']'=4{q_eq_m\over r^2}
\eqno(13b)$$
$$\ro ''=0
\eqno(13c)$$
$$\lapp\po]'=-{m\over r^2}\ro-{q_e^2+q_m^2\over 2r^2}
\eqno(13d)$$
If one requires asymptotic flatness, a solution can be obtained such that
$\po=0$, namely:
$$\fo=-{1\over m}\left({2+q_e^2-q_m^2\over 2r}+{m\over r^2}+{4m^2\over 3r^3}
\right),\qquad\qquad\co=-{2q_eq_m\over mr},$$
$$\ro=-{q_e^2+q_m^2\over 2m}\ ,\qquad\qquad\po=0.
\eqno(14)$$

We pass now to evaluate the second order corrections, which satisfy the
equations:
$$\eqalign{\lap\ft']'=&2{r-2m\over r}\ro\fo '-{48m^2\over r^4}\left(
\fo+{\ro\over r}\right)
+2(q_m^2+q_e^2){\fo\over r^2}\cr&+4(q_m^2-q_e^2){\ro\over r^3}
+2q_eq_m{\co\over r^2}+{r(r-2m)\over 2}\co'^2\cr
\lap\ct']'=&2{r-2m\over r}\ro\co '+
4q_m^2{\co\over r^2}+8q_mq_e\left({\fo\over r^2}-2{\ro\over r^3}\right)\cr
\rt''=&-r\fo'^2+{8m\over r^2}\fo''-{1\over 4}r\co '^2\cr
\lapp\pt]'=&-{r-2m\over 2}\rt ''-{r-m\over r}\rt '
-{m\over r^2}\rt+{2m\over r^3}\ro^2
-(q_e^2-q_m^2){\fo\over r^2}\cr&+2(q_m^2+q_e^2){\ro\over r^3}-
+4m\left({r-2m\over r^2}\fo''-2{r-3m\over r^3}\fo'\right)-q_e q_m{\co\over
r^2}\cr}
\eqno(15)$$
The asymptotically flat solutions of (15), with \bc such that $m$ is the
physical mass of the black hole, are given by:
$$\eqalign{\ft=&-\bigg[{73-45q_m^2\over 60m^3r}+{73-15q_m^2+30q_e^2-15q_m^4-
60q_m^2q_e^2+15q_e^4\over 60m^2r^2}\cr
&+{73+45q_e^2\over 45mr^3}+{73+5q_m^2+50q_e^2\over 30r^4}+{112m\over 75r^5}
+{8m^2\over 9r^6}\bigg]\cr
\ct=&q_eq_m\left[{3\over 2m^3r}+{3-2(q_m^2+q_e^2)\over2m^2r^2}+{2\over 3mr^3}
+{1\over 3 r^4}\right]\cr}
\eqno(16)$$
$$\rt=-\bigg[{4-4(q_m^2-q_e^2)+(q_m^2+q_e^2)^2\over 8m^2r}+{2-q_m^2+q_e^2\over
3mr^2}+{7-3(q_m^2-q_e^2)\over 3r^3}+{16m\over 5r^4}+{24m^2\over 5r^5}\bigg]$$
$$(r-2m)\pt=-\bigg[{1-2(q_m^2-q_e^2)\over 3mr^2}-{11-5(q_m^2-
q_e)^2\over 3r^3}+{(2-50(q_m^2-q_e^2))m\over 15r^4}+{272m^3\over 15r^6}
\bigg]$$
and hence, up to order $\alpha^2$:
$$\eqalign{R^2&\sim r^2-\alpha {q_m^2+q_e^2\over m}r-\alpha^2{1+q_e^2-q_m^2
\over m^2}+O\left({1\over r}\right)\cr
\lambda^2&\sim 1-{2m\over r}-{1+2q_e^2-2q_m^2\over3mr^3}\alpha^2
+O\left({1\over r^4}\right)\cr
\Phi&\sim -\left[{2+q_e^2-q_m^2\over 2m}\alpha+{73-45q_m^2\over 60m^3}
\alpha^2\right]\left({1\over r}\right) +O\left({1\over r^2}\right)\cr
a&\sim -q_eq_m\left[\left({2\alpha\over m}+{3\alpha^2\over m^2}\right){1\over
r}+{5+6q_e^2-2q_m^2\over 2m^2r^2}\alpha^2\right]+O\left({1\over r^2}\right)
\cr}
\eqno(17)$$

The black hole has therefore dilatonic and axionic charge given respectively
at order $\alpha$ by $D={2+q_e^2-q_m^2\over 2m}\alpha$ and $A={2q_eq_m\over m}
\alpha$; these are not however independent parameters, but are functions of
$q_e$, $q_m$ and $m$, in accordance with the weak form of the no-hair
conjecture [2]. The metric has a singularity at $r=r_-\sim\alpha{q_e^2+q_m^2
\over m}$ and a horizon at $r=r_+\sim 2m\left(1-{1-2(q_e^2+q_m^2)^2
\over 12m^2}\alpha^2\right)$. We notice that when $r_->r_+$, the singularity
is naked. This regime is however out of the range of our approximations.

The temperature of the black hole can be readily obtained by requiring the
regularity of the Euclidean section and is given by [4]:
$$T^{-1}\equiv\beta\sim 8\pi m \left[1-\alpha^2\left({\tilde\pt(2m)\over m}-2
\tilde\pt '(2m)\right)\right]=8\pi m \left(1-\alpha^2{73-45(q_m^2-q_e^2)
\over 120m^4}\right)
\eqno(18)$$
This should be compared with the value $T=(8\pi m)^{-1}$ found in [2] when
the order $\alpha '$ corrections to the action are neglected: the temperature
is no longer
independent of the electric and magnetic charge. However, while for a black
hole carrying only magnetic charge the temperature has a maximum and then
goes to zero for a finite value of $m$ [4], the temperature of a purely
electric black hole is at this order of approximation a monotonic decreasing
function of the mass and therefore no massive remnant
is to be expected in the last case.
\bigskip
\centerline{\bf Appendix}

We compare here the perturbative result in the absence of higher order
corrections with the exact solution found in [2].
If one neglects the \GB term, the field equations are, at first order:

$$\eqalign{[r(r-2m)\fo ']'&={q_e^2-q_m^2\over r^2}\cr
\lap\co']'&=4{q_eq_m\over r^2}\cr
\ro ''&=0\cr
\lapp\po]'&=-{m\over r^2}\ro-{q_e^2+q_m^2\over 2r^2}\cr}
\eqno(A.1)$$
whose solutions are:
$$\fo={q_m^2-q_e^2\over 2mr},\qquad\qquad\co=-{q_eq_m\over mr},\qquad\qquad
\ro=-{q_e^2+q_m^2\over 2m},\qquad\qquad\po=0
\eqno(A.2)$$
At second order, the field equations become:
$$\eqalign{\lap\ft']'=&2{r-2m\over r}\ro\fo '
+2(q_m^2+q_e^2){\fo\over r^2}+4(q_m^2-q_e^2){\ro\over r^3}\cr
&+2q_eq_m{\co\over r^2}+{1\over 2}r(r-2m)\co'^2\cr
\lap\ct']'=&2{r-2m\over r}\ro\co '+
4q_m^2{\co\over r^2}+8q_mq_e\left({\fo\over r^2}-2{\ro\over r^3}\right)\cr
\rt''=&-r\fo'^2-{1\over 4}r\chi'^2\cr
\lapp\pt]'=&-{r-2m\over 2}\rt ''-{r-m\over r}\rt '
-{m\over r^2}\rt+{2m\over r^3}\ro^2\cr
&-(q_e^2-q_m^2){\fo\over r^2}+2(q_m^2+q_e^2){\ro\over r^3}\cr}
\eqno(A.3)$$
After a lengthy but straightforward calculation, one obtains:
$$\ft={q_e^4-4q_e^2q_m^2-q_m^4\over 4m^2r^2}\ ,\qquad\ct=-q_eq_m{q_e^2+q_m^2
\over m^2r^2}$$
$$\qquad\rt=-{(q_e^2+q_m^2)^2\over 8m^2r}\ ,\qquad\pt=0
\eqno(A.4)$$
Substituting the results in (10, 11) yields:
$$\lambda^2=1-{2m\over r}\qquad R^2=r^2-\alpha{(q_e^2+q_m^2)\over m}r
\qquad F_{01}={q_e\over r^2}$$
$$\Phi=-\left(\alpha{q_e^2-q_m^2\over 2mr}+\alpha^2{q_e^4-4q_e^2q_m^2-q_m^4
\over 4m^2r^2}\right)+\dots\qquad a=-\alpha{2q_eq_m\over mr^2}\left(1+\alpha
{3q_e^2-q_m^2\over 2mr}\right)+\dots$$
This results coincide with the expansion in $\alpha$ of the exact solution
found in (3):
$$ds^2=-\left(1-{2m\over r}\right)dt^2+\left(1-{2m\over r}\right)^{-1}dr^2
+r\left(r-{\alpha (q_e^2+q_m^2)\over 2m}\right)d\Omega^2$$
$$\Phi=-{1\over 2}\ln\left({mr(mr-\alpha(q_e^2-q_m^2))\over (mr)^2-2\alpha
q_e^2mr+\alpha^2 q_e^2(q_e^2+q_m^2)}\right)$$
$$a=-q_eq_m{2mr-\alpha(q_e^2+q_m^2)\over(mr)^2-2\alpha q_e^2mr+
\alpha^2q_e^2(q_e^2+q_m^2)}$$
$$F_{01}={q_e\over r^2}\qquad\qquad F_{23}={q_m\over r(r-\alpha(q_e+q_m)/2m)}
$$

\def\PRD{Phys. Rev. D}\def\NPB{Nucl. Phys. B}
\def\MPLA{Mod. Phys. Lett. A}

\bigskip
\centerline{\bf Acknowledgements}
I wish to thank N.R. Stewart for many helpful discussions.

\bigskip
\centerline{\bf References}
\halign{#\hfil&\ #\hfil\cr
[1]& D. Garfinkle, G.T. Horowitz and A. Strominger, \PRD 43, 3140 (1991)\cr
[2]& A. Shapere, S. Trivedi and F. Wilczek, \MPLA 6, 2677 (1991) \cr
[3]& A. Sen, {\sl "Electric-magnetic duality in string theory"}, preprint
hep-th 9207053\cr
[4]& S. Mignemi and N.R. Stewart, {\sl "Charged black holes in effective string
theory"}, preprint\cr& hep-th 9212146\cr
[5]& J.H. Schwarz, {\sl "Dilaton-axion symmetry"}, preprint hep-th 9209125\cr
[6]& B.A. Campbell, M.J. Duncan, N. Kaloper and K.A. Olive, \NPB 351, 778
(1991)\cr}
\end